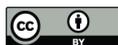

# Observations of planetary waves in the mesosphere-lower thermosphere during stratospheric warming events

N. H. Stray[1], Y. J. Orsolini[2,3], P. J. Espy[1,3], V. Limpasuvan[4], and R. E. Hibbins[1,3]

[1]Department of Physics, NTNU, Trondheim, Norway
[2]Norwegian Institute for Air Research, Kjeller, Norway
[3]Birkeland Centre for Space Science, Bergen, Norway
[4]School of Coastal and Marine Systems Science, Coastal Carolina University, South Carolina, USA

*Correspondence to:* N. H. Stray (nora.kleinknecht@ntnu.no)

**Abstract.** This study investigates the effect of stratospheric sudden warmings (SSWs) on planetary wave (PW) activity in the mesosphere–lower thermosphere (MLT). PW activity near 95 km is derived from meteor wind data using a chain of eight SuperDARN radars at high northern latitudes that span longitudes from 150° W to 25° E and latitudes from 51 to 66° N. Zonal wave number 1 and 2 components were extracted from the meridional wind for the years 2000–2008. The observed wintertime PW activity shows common features associated with the stratospheric wind reversals and the accompanying stratospheric warming events. Onset dates for seven SSW events accompanied by an elevated stratopause (ES) were identified during this time period using the Specified Dynamics Whole Atmosphere Community Climate Model (SD-WACCM). For the seven events, a significant enhancement in wave number 1 and 2 PW amplitudes near 95 km was found to occur after the wind reversed at 50 km, with amplitudes maximizing approximately 5 days after the onset of the wind reversal. This PW enhancement in the MLT after the event was confirmed using SD-WACCM. When all cases of polar cap wind reversals at 50 km were considered, a significant, albeit moderate, correlation of 0.4 was found between PW amplitudes near 95 km and westward polar-cap stratospheric winds at 50 km, with the maximum correlation occurring ∼ 3 days after the maximum westward wind. These results indicate that the enhancement of PW amplitudes near 95 km is a common feature of SSWs irrespective of the strength of the wind reversal.

## 1 Introduction

Stratospheric sudden warmings (SSWs) are dramatic breakdowns of the polar vortex occurring in the polar wintertime that can dynamically couple the atmosphere all the way from the troposphere into the ionosphere (e.g., Limpasuvan et al., 2004; Goncharenko et al., 2010; Pancheva and Mukhtarov, 2011; Yuan et al., 2012). They occur frequently in the northern polar hemisphere but vary in strength and time of occurrence. SSWs are caused by the interaction of planetary waves (PWs) with the mean flow (Matsuno, 1971) that leads to an abrupt reversal of the zonal-mean winds in the middle atmosphere as well as to a sudden warming (cooling) of the stratosphere (mesosphere) (e.g., Manney et al., 2008; Chandran et al., 2014, and references therein).

A total breakdown of the polar vortex during an SSW can lead to a nearly isothermal region around the stratopause, with the stratopause subsequently re-forming at higher altitudes (Manney et al., 2008). Such an event is known as an elevated stratopause (ES) event. Based on a case study with the WACCM model, Limpasuvan et al. (2012) suggested that during a strong warming accompanied by an ES event, PWs appear in the mesosphere–lower thermosphere (MLT). Along with westward gravity wave drag, momentum deposition by PWs also contributes to the formation and descent of the elevated stratopause (Limpasuvan et al., 2012). The sudden changes in atmospheric conditions related to an SSW alter the transmission of planetary and gravity waves and result in large vertical and horizontal temperature and velocity gradients that can lead to the generation of new waves (Chandran et al., 2013b). In a high-resolution model study, Tomikawa et al. (2012) demonstrated that the generation of the MLT planetary waves could arise from large-scale flow instabilities. In a climatological analysis using both the WACCM model and the MERRA analysis data, Chandran et al. (2013a) found that over half of the SSW events were accompanied by an ES, and ∼ 87 % of these combined events showed enhanced planetary wave activity in the upper mesosphere.

In addition to these model results, Chandran et al. (2013b) also observed an enhancement of PW amplitudes and a change of their longitudinal propagation speed in the MLT in connection with a strong SSW event in January 2012 using the Specified Dynamics Whole Atmosphere Community Climate Model (SD-WACCM) and temperatures from the Sounding of the Atmosphere using Broadband Emission Radiometry (SABER) instrument. This event was noteworthy in that it was strong, with the total zonal-mean zonal wind at 50 km (averaged from 50–77° N) reaching westward wind speeds of more than 40 m s$^{-1}$ during the reversal. In addition it was accompanied by an elevated stratopause event. Chandran et al. (2013b) inferred that the PWs in the MLT were generated in situ by the instabilities associated with the large wind and temperature gradients that resulted from this strong SSW event.

Here we examine whether PW enhancements in the MLT are a general feature connected to SSWs or whether they are only associated with strong events. To do so, PW activity around 95 km was examined during several SSW events that were characterized by a range of wind reversal magnitudes (−2 to −50 m s$^{-1}$). The PW amplitudes were derived using the meridional meteor winds from a northern high-latitude chain of SuperDARN meteor radars for the winter seasons between 2000 and 2008.

## 2 Data

PW activity near 95 km has been derived from meteor winds measured by a chain of SuperDARN radars at high northern latitudes (51–66° N). SD-WACCM has been used to model PW activity throughout the middle atmosphere and monitor atmospheric background conditions of wind and temperature.

### 2.1 Observational data (SuperDARN)

Hourly meridional meteor winds from eight SuperDARN radars (Greenwald et al., 1985, 1995) located between 51–66° N and 150° W–25° E have been used to extract the longitudinal structure of PWs with zonal wave number 1 and 2 ($S_1$ and $S_2$) in the MLT, at approximately 95 km (Hall et al., 1997; Hibbins and Jarvis, 2008) for all winters (November–March) between January 2000 and December 2008. This has been done following the technique described and validated in Kleinknecht et al. (2014). Briefly, after an initial quality check, daily mean winds are produced by fitting and removing tidal (8, 12, 24 h) and 2-day wave components to 4-day segments of the hourly winds which are stepped in 1-day intervals. These resulting daily winds are then fitted as a function of longitude to provide the amplitude and phase of the $S_1$ and $S_2$ PWs. Kleinknecht et al. (2014) verified the amplitude and phase of the retrieved wave number components to correlate well (correlation coefficient: 0.9) with an ideal fit covering 360° of longitude (2.5° longitude spacing). The amplitude of the $S_1$ and $S_2$ wave components agreed with the ideal fit within the fitting uncertainties (±20 and ±10 %, respectively). The wave number 1 ($S_1$) and 2 ($S_2$) components resulting from the fit to the chain of SuperDARN radars are shown in Fig. 1 for all winter seasons (November–March) between 2000 and 2008. The independent fits from the individual days are presented in the form of a Hovmöller diagram, where the phase vs. longitude of the $S_1$ and $S_2$ components can be seen to evolve in time. Red and blue colors signify poleward and equatorward wind, respectively. Each wave component represents the superposition of all stationary as well as eastward and westward traveling waves with that zonal wave number. The temporal changes in the amplitude and longitudinal phase of each wave component, shown in the Hovmöller diagram, indicate the interaction of the different temporal components of the PW as they propagate in the zonal background wind.

### 2.2 Model data (SD-WACCM)

WACCM is a global circulation model of the National Center of Atmospheric Research (NCAR) extending from the surface to 150 km (88 pressure levels) with fully coupled chemistry and dynamics. Its horizontal resolution is 1.9° × 2.5° (latitude × longitude). The WACCM version used in this study is a specified dynamics version of WACCM4 called SD-WACCM. The dynamics and temperature of the specified dynamics version are nudged to MERRA, the Modern-Era Retrospective Analysis for Research and Application of the NASA Global Modeling and Assimilation Office (Rienecker et al., 2011) up to 0.79 hPa (∼ 50 km). Meteorological fields above this altitude are fully interactive with linear transition in between. Due to the nudging SD-WACCM is capable of representing atmospheric temperature and dynamics for individual dates and has been shown to represent ES events well (Eyring et al., 2010; Chandran et al., 2013a; Tweedy et al., 2013).

## 3 Data analysis

SSWs occur frequently in the Northern Hemisphere during winter (November–March). However their time of occurrence and their strength vary due to differences in the stratospheric wave forcing. To investigate the effect of SSWs on PW activity in the MLT, a threshold to define an SSW event has to be set. Here, the onset day of an event was defined as the point when the zonal-mean zonal polar cap (70–90° N) wind at 0.7 hPa (∼ 50 km) reversed from eastward to westward and persisted reversed for at least 4 days. The reversal of the polar cap wind at 50 km was used to identify the events following Tweedy et al. (2013), who found that this criterion was a better indicator of the wind reversal extending into the mesosphere and the onset of vertical upwelling than the WMO (World Meteorological Organization) definition of SSWs at 10 hPa and 60° latitude. Using the polar cap wind reversal at 50 km, only two of the events classified as strong would be considered to be major stratospheric warmings according to the WMO definition. SSW events have been defined as strong events when their wind reversal in the stratospheric (50 km) polar cap wind (70–90°) persists for 4 or more days and exceeds westward wind magnitudes of $10\,\mathrm{m\,s^{-1}}$. Zonal wind data from MERRA were used to identify SSW events in the zonal-mean polar cap zonal winds. In addition SSWs were divided into SSWs with and without elevated stratopause (ES) events. An ES event was defined as having a temperature below 185 K at 80 km immediately after the wind reversal, and a stratopause elevation of at least 10 km after the onset of the wind reversal. Here SD-WACCM was used to define the events due to the limited top altitude of MERRA. Using these criteria, in total 23 SSW events have been found during the years 2000–2008, among which 7 of these events were followed by an ES event (11 December 2000, 29 January 2001, 22 December 2001, 29 December 2002, 19 December 2003, 9 January 2006, 23 January 2008). It should be noted that no polar cap wind reversals lasting at least 4 days occurred 40 days prior to the seven SSW ES events studied, ensuring that the baseline was not disturbed by a previous event. Composites of these seven SSW ES events were made for the modeled temperature, wind and PW activity using SD-WACCM and for the PW activity observed by the SuperDARN chain.

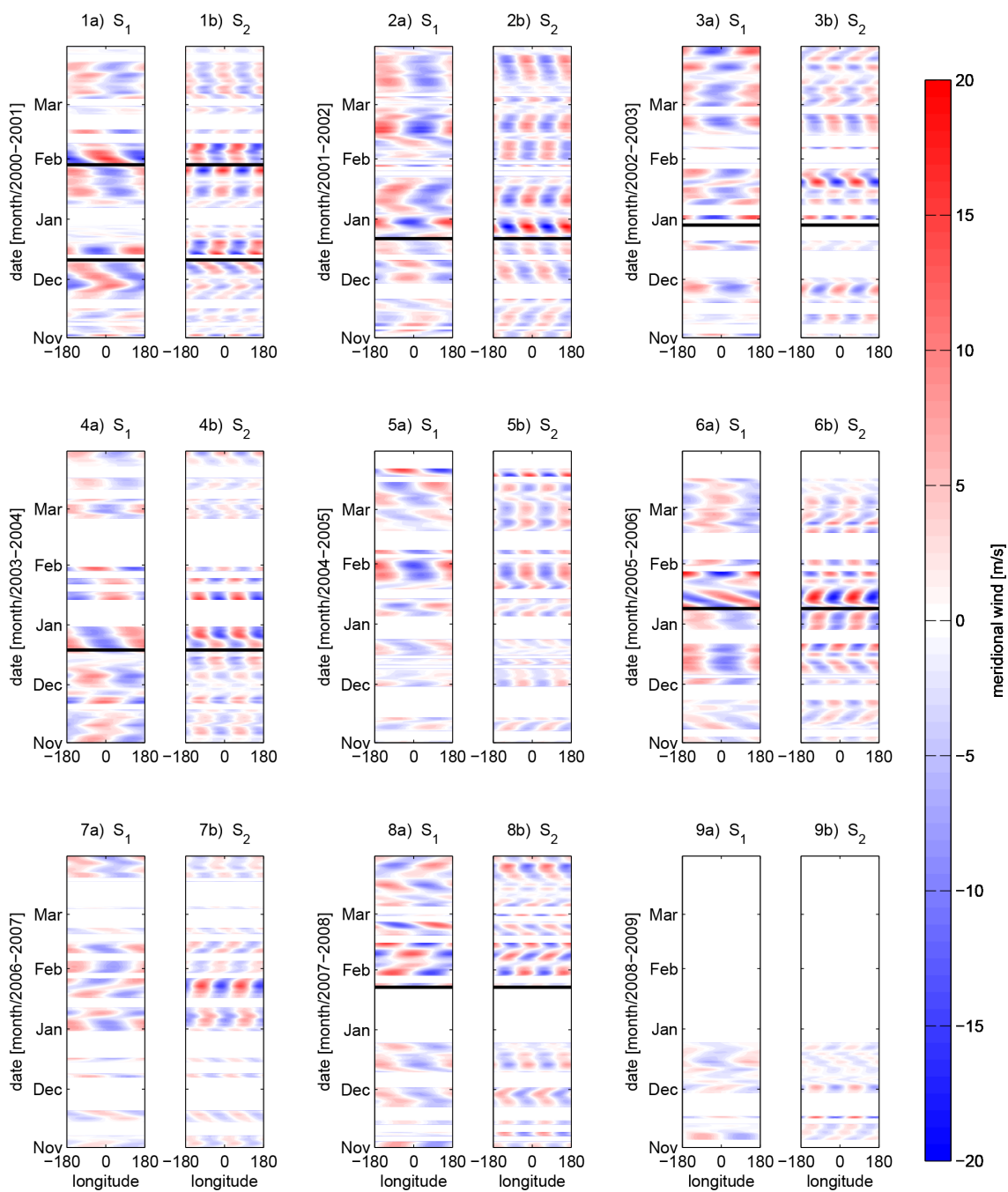

**Figure 1.** Longitudinal wave components $S_1$ **(a)** and $S_2$ **(b)** of the mean meridional wind anomalies for winters 2000/2001–2008/2009 (1–9). Red and blue colors signify poleward and equatorward wind, respectively. The horizontal black lines indicate the onset of the wind reversal for each of the SSW events that were accompanied by an elevated stratopause (see Sect. 3).

## 4 Modeled wind and temperature during a composite SSW ES event

The composite of the SD-WACCM background conditions (temperature and wind) using these seven SSW ES events is shown in Fig. 2 and demonstrates the general behavior of the atmospheric zonal-mean temperature and winds during an SSW ES event. The upper panel depicts the zonal-mean zonal wind. Red and blue colors signify eastward and westward winds, respectively. The lower panel shows the zonal-mean temperature with red and blue colors signifying temperatures above and below 220 K, respectively.

The first days of the composite fields reflect undisturbed winter conditions. Briefly, the zonal-mean zonal wind is eastward in the stratosphere, which leads to filtering of the eastward-propagating gravity waves and hence westward gravity wave momentum deposition and westward zonal-mean winds in the mesosphere. The vertical temperature gradient is positive (increasing with altitude) in the stratosphere and negative in the mesosphere, with the stratopause clearly visible as a temperature maximum around 60 km.

An SSW is created by the interaction of PWs with the mean flow that leads to a deceleration of the polar vortex and hence a reversal of the zonal-mean zonal wind (Matsuno, 1971). In Fig. 2a this reversal of the zonal wind can be seen as a strong, westward wind regime in the stratosphere starting at day zero of the composite and lasting for several days. This distortion of the polar vortex leads to sinking motion and hence to adiabatic heating in the stratosphere, and a descent of the stratopause as can be seen in Fig. 2b. In addition the distortion of the polar vortex leads to a change in gravity wave filtering. During the wind reversal eastward gravity waves can reach the mesosphere while westward gravity waves are blocked (de Wit et al., 2014). This leads to the deposition of eastward momentum and a reversal of the zonal mean wind in the mesosphere as can be seen in Fig. 2a. In addition, there is an accompanying reversal of the meridional circulation and hence a rising motion and cooling in the mesosphere (e.g., Limpasuvan et al., 2012; Chandran et al., 2014, and references within), as can be seen in Fig. 2b. When an SSW is accompanied by an ES event, the stratopause disappears temporarily and then re-forms at higher altitudes (Fig. 2b). The disappearance of the stratopause is related to a complete breakdown of the polar vortex and a nearly isothermal region between the stratosphere and the mesosphere. The stratopause can then reform at higher altitudes (∼ 70–80 km) as a so-called elevated stratopause (e.g., Manney et al., 2008). Although each of the seven events shows a stratopause jump between 20 and 42 km, the composite shown in Fig. 2 is indexed to zero on the day of the wind reversal at 50 km (defined to be the onset date) and not at the occurrence of the ES event. Therefore the mean elevated stratopause, although clearly visible at around 70 km after the warming, is not representative of the individual stratopause jumps in the composite.

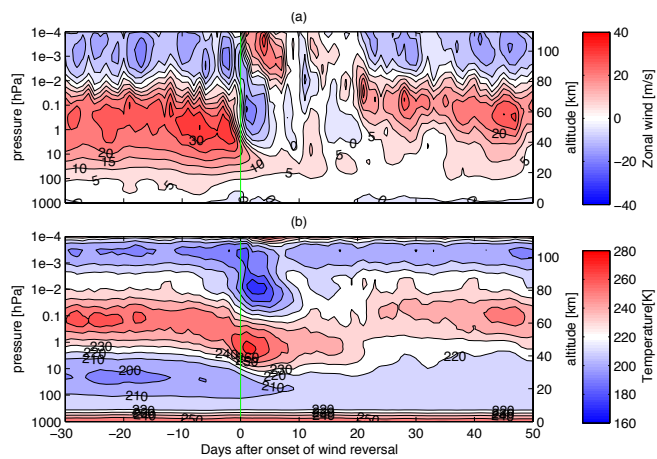

**Figure 2.** Composite (based on seven SSW ES events) of polar cap (70–90°) zonal-mean zonal wind **(a)** and temperature **(b)** observed by SD-WACCM. Day zero (vertical green line) marks the onset of the polar cap zonal-mean zonal wind reversal at 0.7 hPa (∼ 50 km).

## 5 Results

### 5.1 MLT planetary wave during SSW ES events

The Hovmöller diagrams presented in Fig. 1 also show the onset of the stratospheric wind reversal for each of the seven SSW events that were accompanied by an elevated stratopause, as marked by the horizontal black lines. The wind reversals triggering those events cover maximum magnitudes between $-11$ and $-30\,\mathrm{m\,s^{-1}}$ and last for 4 to 16 days. However the observations show consistent PW behavior during SSW events. That is, each event is accompanied by an increase in PW amplitude. The PW activity after the onset of the wind reversal can be observed to propagate with quite stable phase speed for approximately 10 days. The seven SSW ES events investigated here reveal phase speeds between $-45$ and $15°$ longitude per day. Three events show the phase speed to be stronger westward, and one event shows eastward phase speed after the reversal. For the other three events the phase speed is not significantly different before and after the reversal.

We also observe a coherent PW response during an SSW ES event by forming a composite of PW activity for the seven SSW ES events. The composite of the PW activity observed by the SuperDARN radar chain is presented in Fig. 3. It shows the PW amplitudes in the MLT ($\sim 95\,\mathrm{km}$) for the $S_1$ (a), the $S_2$ (b) and the sum of both zonal wave components (c). In addition the stratospheric ($\sim 50\,\mathrm{km}$) polar cap zonal-mean zonal wind is plotted in magenta, and the onset of the stratospheric wind reversal, day zero of the composite, is marked with a vertical green line. A significant increase in PW activity can be seen just after the stratospheric zonal-mean zonal wind reversal in both wave numbers. The enhancement occurs slightly earlier in the $S_1$ component but is stronger in the $S_2$ component. It should be noted that all the seven events used for the composite happen to be associated with a vortex displacement.

For comparison with the SuperDARN observations, Fig. 4 shows the composite of the modeled PW activity, derived from the meridional wind of SD-WACCM, from the ground to $10^{-4}$ hPa ($\sim 110\,\mathrm{km}$) for the same seven SSW ES events. The zonal wave number 1 component ($S_1$) is shown in the upper panel, the zonal wave number 2 component ($S_2$) in the middle panel and the sum of both wave components in the lower panel. The vertical green line marks the onset of the stratospheric wind reversal. The horizontal blue line marks the approximate mean altitude of the SuperDARN wind observations. As expected, strong PW activity can be seen in the stratosphere leading to the wind reversal and the SSW. In the model, the PW activity minimizes at around 80 km due to either strong gravity wave drag (e.g., Smith, 2003) and/or a strong negative wind shear (e.g., Smith, 1983; McDonald et al., 2011). However, similar to the observed PW activity, there is an enhancement in PW activity in the MLT (above 80 km) just after the onset of the stratospheric wind reversal. While stratospheric PW activity seems to be dominated by wave number 1, the amplitudes of the two zonal components ($S_1$ and $S_2$) are quite similar in the MLT, with the amplitudes of the $S_2$ component being slightly stronger. Similar to the PW observed in the MLT, the modeled amplitudes of the $S_1$ component peak slightly before the $S_2$ component, although the modeled peaks occur slightly earlier than in the observations.

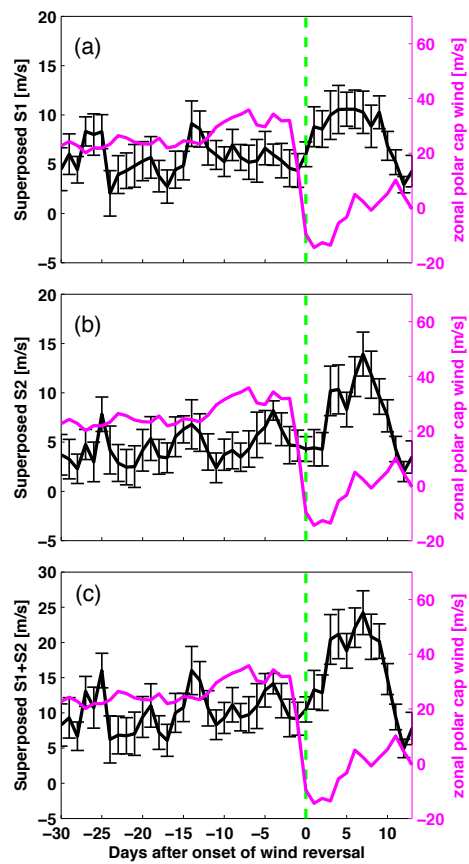

**Figure 3.** Composites of PW activity observed by SuperDARN around SSW events for the $S_1$ **(a)**, $S_2$ **(b)** and $S_1 + S_2$ **(c)** components (black) and the stratospheric ($\sim 50\,\mathrm{km}$) polar cap zonal-mean zonal wind (magenta). The onset of the wind reversal is marked with a vertical green line at day zero.

In summary, the composites of the SSW events that are followed by an elevated stratopause in both the observed (Fig. 3) and the modeled (Fig. 4) cases show a clear increase in PW amplitude above 80 km after the onset of the SSW. In addition a shift of the propagation direction often occurs after the onset of the SSW (Fig. 1). This indicates that SSWs accompanied by an ES event in general have a strong influence on PW activity in the MLT, irrespective of the strength of the reversal.

## 5.2 MLT planetary wave activity during stratospheric wind reversals

The previous sections showed the typical behavior of winds, temperatures and PWs during the seven SSWs accompanied by ES events irrespective of the strength of the reversal. In this section the correlation between all stratospheric wind reversals and MLT planetary wave activity is investigated. This includes all 23 SSW events and also 6 additional events where wind reversal is shorter than 4 days. Figure 5 shows the PW amplitudes ($S_1 + S_2$) observed by the Super-

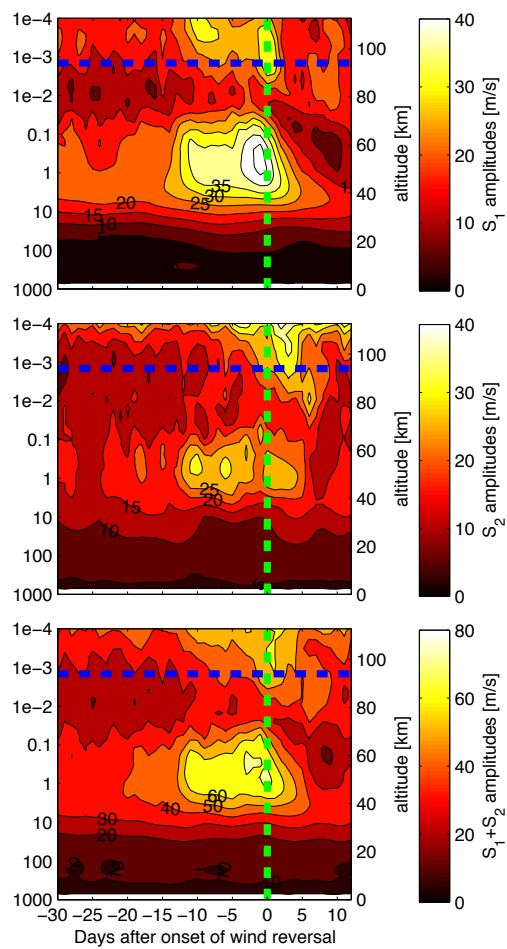

**Figure 4.** Composites of PW activity observed by SD-WACCM around SSW events for the $S_1$ (upper panel), the $S_2$ (middle panel) component and the sum of both components (lower panel) averaged over 48–65° N. The vertical green line marks the onset of the stratospheric wind reversal. The horizontal blue line marks the approximate mean altitude of the SuperDARN wind observations.

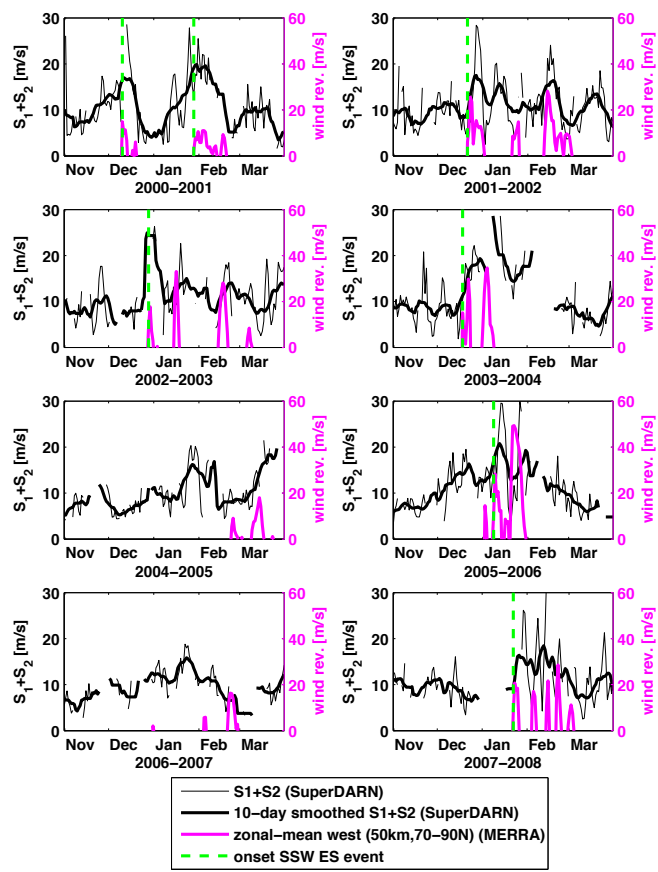

**Figure 5.** MLT planetary wave amplitudes from SuperDARN (black) and stratospheric westward polar cap (70–90° N) winds at 0.7 hPa (∼ 50 km) from MERRA (magenta) during winters 2000/2001–2007/2008. The PW amplitudes (black), the sum (thin black line) and the 10-day smoothed sum (thick black line) of the wave number 1 and 2 components ($S_1 + S_2$) retrieved from the SuperDARN data. The vertical green line marks the onset of the stratospheric wind reversal of the composited events.

DARN radar chain (black) for all winters between 2000/2001 and 2007/2008 together with the westward component of the polar cap wind at 50 km retrieved from MERRA (magenta). The magnitudes of the reversals vary between −2 and −50 m s$^{-1}$ and last between 1 and 19 days. The SSWs used for the composite study in the previous section are marked with vertical green lines.

A similar composite analysis to that presented in Sect. 5.1 using all stratospheric wind reversals could not be used to investigate the relationship between the wind reversal and PW amplitudes in the MLT because many of them occur shortly after an SSW ES event that has perturbed the background conditions. Therefore, to investigate the general correlation between stratospheric wind reversals of variable strength and the MLT PW amplitudes, polar cap westward winds at 50 km (MERRA) and the observed PW amplitudes (SuperDARN) have been correlated between November and March.

Only days for which both the westward polar cap winds and PW measurements exist were correlated. The correlation is shown in Fig. 6. The highest correlation (correlation coefficient = 0.4) occurred with the westward polar cap wind maximum leading the PW activity maximum by 3 days. The correlation between the westward wind and PW bursts is only moderate but more than 99 % significant and shows that MLT PW enhancements are not just associated with SSWs that are accompanied by ES events, but they are a general feature attendant with stratospheric polar cap wind reversals of any strength. To make sure that the observed correlation is not being triggered by the SSW ES events but by the bulk of wind reversals, the correlation has been repeated excluding the SSW ES events used in the previous composite study. The correlation coefficient (not shown) becomes slightly smaller (0.3) but still peaks at a 3-day lag (polar cap winds leading) and remains more than 99 % significant.

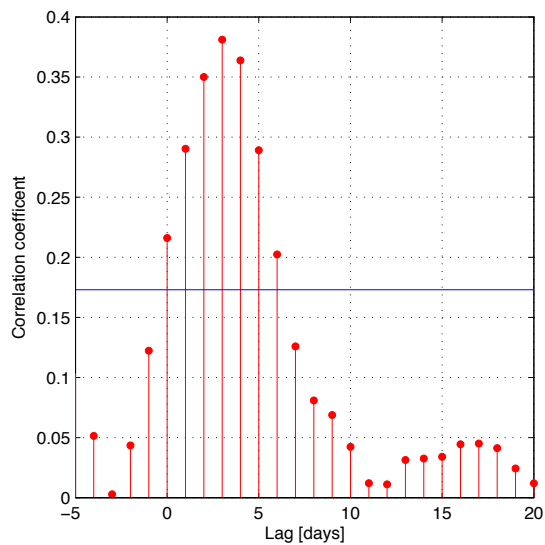

**Figure 6.** Correlation between westward polar cap stratospheric winds ($\sim 50$ km) from MERRA and PW amplitudes ($S_1 + S_2$) from MLT meridional SuperDARN winds ($\sim 95$ km). The stratospheric westward wind leads the PW enhancements in the MLT by approximately 3 days with a moderate correlation coefficient of 0.4. The horizontal blue line represents the 99 % confidence level.

## 6 Discussion and conclusion

Limpasuvan et al. (2012) mentioned in a case study using WACCM that the ES event was accompanied by a PW in the MLT region, and Chandran et al. (2013a) later established a climatology of SSW events using WACCM. Chandran et al. (2013a) reported that the majority of SSW events studied were accompanied by an increase in PW amplitudes in the MLT. In addition, Chandran et al. (2013b) observed such an increase in PW amplitudes, using SABER data, in a single, strong (polar cap zonal wind reversal at 50 km greater than $-40\,\mathrm{m\,s^{-1}}$ (max westward wind)) SSW ES event in January 2012. Our observations (Fig. 1) indicate the enhancement in PW amplitude approximately 5 days after the wind reversal (Fig. 3) to be a consistent feature of all seven SSW ES events, irrespective of the magnitude ($-11$ to $-30\,\mathrm{m\,s^{-1}}$) and duration (4 to 16 days) of the wind reversal. The consistent increase in planetary wave amplitudes near 95 km observed in the SuperDARN wind data that occurs after the onset of the SSW ES events (shown by the horizontal black lines) can be observed in the Hovmöller diagrams of Fig. 1 as well as in the temporal evolution of the PW amplitudes and winds shown in Fig. 3. In addition these observations are consistent with the modeled increase of PW amplitude in the MLT shown in Fig. 4. Furthermore the relation between the observed PW activity and all wind reversals (Fig. 6), including a variety of polar cap zonal wind reversals between $-2$ and $-50\,\mathrm{m\,s^{-1}}$ that last between 1 and 19 days, is striking. Their moderate but significant correlation indicates PW bursts in the MLT to lag the maximum of the stratospheric wind reversals by 3 days. This 3-day lag between the maximum wind reversal and the maximum PW activity observed in the correlation analysis (Fig. 6) is consistent with the timing observed in the composite analysis (Fig. 3) which shows the maximization of PW activity approximately 5 days after the onset of the reversal and hence approximately 3 days after the maximum of the wind reversal.

Chandran et al. (2013b) observed not only an increase in PW amplitude after the 2012 SSW ES event but also a change of the PW longitudinal phase speed towards stronger westward propagation and the occurrence of 5–10-day westward-propagating waves following the onset of the event. Here we observe an increase in westward propagation in many of the SSW ES events. However, during some SSW ES events no clear phase shift and even eastward wave propagation can be observed after the onset of the SSW.

All these observations together indicate that the enhancement of PW activity in the MLT is a general feature connected to SSWs in the Northern Hemisphere and not just related to strong events, confirming the modeling results of Chandran et al. (2013a). Indications on how the PW in the MLT is related to the SSW can be collected from the modeled background conditions and the modeled PW activity presented in Figs. 2 and 4, respectively. While the enhancement in the observed PW amplitudes shown in Fig. 2 occurs during times of zonal-mean eastward winds in the MLT, the wind below this region is strongly westward during the SSW event and would inhibit upward-propagating PW into this favorable wind regime. Indeed, the modeled SD-WACCM PW activity clearly shows that the amplitudes of PWs propagating up from the stratosphere minimize when they reach altitudes around 80 km before building again in the MLT. This minimum could be related to ducting of PWs (Limpasuvan et al., 2012) or indicate that the PW activity in the MLT observed after the onset of the stratospheric wind reversal might be locally generated and is not just a continuation of the stratospheric PW activity. Such in situ generation of secondary PW activity in the MLT due to zonal asymmetry imposed by gravity wave drag during an SSW has been suggested by, for example, Liu and Roble (2002). Furthermore the model results presented above (Fig. 2) show very strong temperature and wind gradients to develop during all observed SSW events which favor the development of baroclinic and barotropic instabilities (e.g., Matsuno, 1971; Pedlosky, 1979; Nielsen et al., 2010; Limpasuvan et al., 2012; Tomikawa et al., 2012). This suggests that, like in the strong SSW event studied by Chandran et al. (2013b), instabilities associated with those gradients are potentially an additional trigger for the enhanced PW activity seen in the MLT during stratospheric ($\sim 50$ km) polar cap wind reversals.

*Acknowledgements.* The authors acknowledge the use of Super-DARN data. SuperDARN is a collection of radars funded by the national scientific funding agencies of Australia, Canada, China, France, Japan, South Africa, United Kingdom and United States of America. This study was partly supported by the Research Council of Norway/CoE under contract 223252/F50. V. Limpasuvan was supported by the National Science Foundation under grant AGS-1116123 and the Coastal Carolina University Kerns Palmetto Professorship. The authors are grateful for the help of D. E. Kinnison at NCAR for his assistance with SD-WACCM.

Edited by: Q. Errera